\documentstyle[12pt, amsfonts]{article}

\def\theequation{\thesection.\arabic{equation}}
\def\be{\begin{equation}}
\def\ee{\end{equation}}
\def\ba{\begin{eqnarray}}
\def\ea{\end{eqnarray}}
\def\lb{\label}

\def\a{\alpha}
\def\b{\beta}
\def\d{\delta}

\def\bq{\overline{q}}

\def\Z{\Bbb Z}

\def\id{\mbox{\rm 1\hspace{-3.5pt}I}}

\def\R{\hat{R}}
\def\Rp{\hat{R}(p)}
\newcommand{\DR}[1]{\hat{R}_{#1}(p)}

\def\eu2{\varepsilon^{|2  \dots n{+}1{\cal i}}}
\def\ed2{\varepsilon_{{\cal h}2  \dots n{+}1|}}
\def\eu2p{\varepsilon^{|2  \dots n{+}1{\cal i}}(p)}
\def\ed2p{\varepsilon_{{\cal h}2  \dots n{+}1|}(p)}

\def\U2{ {\cal E}^{|2  \dots n{+}1{\cal i} } }
\def\D2{ {\cal E}_{{\cal h}2  \dots n{+}1| } }
\def\Eu2p{{\cal E}^{|2  \dots n{+}1{\cal i}}(p)}
\def\Ed2p{{\cal E}_{{\cal h}2  \dots n{+}1|}(p)}

\def\subbbc{{\rm C}\kern-3.3pt\hbox{\vrule height4.8pt width0.4pt}\,}

\normalbaselines
\begin{document}
\begin{center}
\vspace*{1.0cm}

{\LARGE{\bf A Finite Dimensional Gauge Problem in the WZNW
Model
\footnote[0]{Talk presented by I.T. Todorov at the International
symposium "Quantum Theory and Symmetries", 18-22 July 1999, Goslar, Germany}
}}

\vskip 1.5cm

{\large{\bf
M. Dubois-Violette$^{1,7}$,
P. Furlan$^{2,3}$,
L.K. Hadjiivanov$^{4,5}$,
A.P.~Isaev$^6$,
P.N. Pyatov$^6$,
I.T. Todorov$^{4,7}$ }}
\vskip 0.5 cm
$^1$Laboratoire de Physique Th\'eorique, B\^at. 210,
Universit\'e Paris XI,\\
F-91405 Orsay Cedex, France\\
$^2$Dipartimento di Fisica Teorica dell' Universit\`a di Trieste,
I-34100 Trieste, Italy\\
$^3$Istituto Nazionale di Fisica Nucleare (INFN),
Sezione di Trieste, Trieste, Italy,\\
$^4$Division of Theoretical Physics, Institute for Nuclear
Research and Nuclear Energy,\\
Bulgarian Academy of Sciences, Tsarigradsko Chaussee
72, BG-1784 Sofia, Bulgaria,\\
$^5$Abdus Salam International Centre for Theoretical Physics (ICTP),\\ 
P.O. Box 586, I-34100 Trieste, Italy\\
$^6$Bogoliubov Laboratory of Theoretical
Physics, JINR, Dubna,\\
141 980 Moscow Region, Russia\\
$^7$Erwin Schr\"odinger Institute for Mathematical
Physics (ESI), A-1090 Wien, Austria

\end{center}
\vspace{0.5 cm}


\begin{abstract}
The left and right zero modes of the
$SU(n)$ WZNW model give rise to a pair of isomorphic
mutually commuting algebras $\cal A$ and $\overline{\cal A}\,.$
Here $\cal A$ is the {\em quantum matrix algebra} \cite{HIOPT}
generated by an $n\times n$ matrix
$a=\left( a^i_\alpha \right)\,,i, \a =1,\dots ,n\,$
(with noncommuting entries) and an abelian group consisting of
products of $n$ elements $q^{p_i}$ satisfying
$\prod\limits_{i=1}^n q^{p_i} = 1,\,\, q^{p_i} a^j_\a =
a^j_\a q^{p_i + \delta^j_i - {1\over n}}\,.$
For an integer ${\widehat{su}}(n)$ {\em height} $h\ (\,=k+n\ge n)\,$
the complex parameter $q\,$ is an even root of unity, $q^h=-1\,,$ and
$\cal A$ admits an ideal ${\cal I}_h$ generated by
$\{ \left( a^i_\alpha\right)^h ,\,\, q^{2p_{ij}h} - 1,\,\,
p_{ij}=p_i-p_j,\,\, \a ,i, j = 1,\ldots , n \}\,$ such that the
factor algebra ${\cal A}_h ={\cal A}/{\cal I}_h$ is
finite dimensional. The structure of superselection sectors of
the (diagonal) $2$-dimensional ($2D$) WZNW
model is then reduced to a finite dimensional problem of a gauge
theory type. For $n=2\,$ this problem is solved using a
generalized BRS formalism.

\vspace{1 cm}
\noindent{L.P.T.-ORSAY 99-73, ESI 778}

\end{abstract}


\vspace{1 cm}

\section*{Introduction}
Although the Wess--Zumino--Novikov--Witten
(WZNW) model is first formulated in terms of a (multivalued)
action \cite{W} , it is originally solved \cite{KZ}
by using axiomatic conformal field theory methods.
The two dimensional (2D) Euclidean Green functions are written
\cite{BPZ} as sums of products of analytic and antianalytic {\em
conformal blocks}. Their operator interpretation exhibits some
puzzling features: the presence of {\em noninteger} ("quantum")
{\em statistical dimensions} (that appear as positive real
solutions of the {\em fusion rules} \cite{V}) contrasted with the
local ("Bose") commutation relations (CR) of the corresponding 2D
fields. The gradual understanding of both the factorization
property and the hidden {\em braid group statistics} (signaled by
the quantum dimensions) only begins with the development of the
canonical approach to the model (for a sample of references, see
\cite{B} - \cite{FHT3}) and the associated splitting of the basic group
valued field $g : \; {\Bbb S}^1 \times {\Bbb R}\;\rightarrow\;
G\;$ into chiral parts. The resulting zero mode extended phase
space displays a new type of {\em quantum} group {\em gauge
symmetry}: on one hand, it is expressed in terms of the {\em
quantum universal enveloping algebra} $U_q({\cal G})\,,$ a
deformation of the {\em finite dimensional Lie algebra} ${\cal
G}\,$ of $G\,$ -- much like a gauge symmetry of the first kind; on
the other, it requires the introduction of an {\em extended,
indefinite metric state space}, a typical feature of a (local)
gauge theory of the second kind.

Chiral fields admit an expansion into {\em chiral vertex
operators} (CVO)
which are characterized essentially by the currents' degrees
of freedom with "zero mode" coefficients that are independent of
the world sheet coordinate \cite{AF, BF, FHT2, FHT3}.
Such a type of quantum theory has been studied in the framework
of lattice current algebras (see \cite{F2, G, FG, AFFS, BS}
and references therein)
and has not been brought to a form yielding a satisfactory
continuum limit . The direct investigation of the quantum model
\cite{FHT2, FHT3, HIOPT} has singled out a nontrivial gauge
theory problem. This problem has been tackled in two steps
\cite{DT, DT2} in terms of a generalization of the
Becchi-Rouet-Stora (BRS) \cite{BRS} cohomologies.

The present paper aims at providing a concise survey
of this study (including a preview of a work in progress
\cite{FHIPT}). It is organized as follows. Section 1 is devoted
to a brief review of results of Gaw\c{e}dzki et al. \cite{G, FG}
that culminate earlier work \cite{B, F1, F2} on the canonical
approach to the classical WZNW model and on the first steps to its
quantization -- including the $R$-matrix exchange relations.
Section 2 defines and studies the main object of interest to us, the
quantum matrix algebra ${\cal A}$. The basic exchange relations
for $a^i_\a \in{\cal A}\,$ involve a dynamical
$R$-matrix $\R (p)\,$ of Hecke type. Section 3 introduces
the Fock space (${\cal F}$) representation of
${\cal A}\,$ equipped with two $U_q(sl_n)\,$ invariant
forms, a bilinear and a sesquilinear one.
Both forms have a
kernel ${\cal I}_h {\cal F}\,$ where ${\cal I}_h\,$ is the maximal
ideal of ${\cal A}\,$ (for $q^h = -1\,$);
the corresponding quotient
of ${\cal F}_h\,$ with respect to this kernel
is finite dimensional. Section 4
identifies "the physical state space" for the $2D\,$ zero mode
algebra ${\cal A}\otimes{\overline{\cal A}}\,$
in terms of a generalized BRS
cohomology (with a BRS charge $Q\,$ such that $Q^h = 0\,, \
h>n\,$). In the concluding remarks (Section 5) we indicate some
open problems.

\section{Canonical approach to the WZNW model}
The starting point of the (first order) covariant Hamiltonian 
formulation of the WZNW model is the canonical 3-form \cite{G}
\be
\lb{can3}
\omega = d\a -{k\over{12\pi}}\; {\rm tr}\; (g^{-1} dg )^3\,,\quad \a =
{1\over 2}\; {\rm tr}\; \left( i (j^0 dx - j^1 dt)\; g^{-1} dg
- {{\pi}\over k} j_\nu j^\nu\; dt\; dx\right)\ \
\ee
$(j_\nu j^\nu = j_1^2 - j_0^2\, )\,.$
The coefficient to the WZ term (the closed but not exact form
${\rm tr}\; (g^{-1} dg )^3$) is chosen in such a way that the solution of
the resulting equations of motion (cf. \cite{G, FHT1})
\be
\lb{eqmo}
j_\nu = {k\over{2\pi i}}g^{-1} \partial_\nu g\ ,\quad \partial_\nu
j^\nu = {{2\pi i}\over k} [j_0 , j_1 ] 
\ee
splits into left and right movers. Indeed, Eqs.(\ref{eqmo}) imply
\be
\lb{cone}
\partial_+ (g^{-1} \partial_- g) = 0\ \Leftrightarrow\ 
\partial_- ( (\partial_+ g)g^{-1} ) = 0\,,\quad 2\partial_\pm =
\partial_1 \pm \partial_0\quad (\partial_\nu =
{{\partial}\over{\partial x^\nu}}\,, \ x^0 \equiv t )\,.
\ee
The general solution of Eq.(\ref{cone}) is given by
\be
\lb{g}
g^A_B (t, x) = u^A_\a (x+t)\, ({\bar u}^{-1})^\a_B (x-t)\quad
({\rm classically}\ g, u, {\bar u} \in G=SU(n) )
\ee
provided the periodicity condition for $g\,$ is weakened to a
twisted periodicity condition for $u\,$ and ${\bar u}\,,$
\be
\lb{per}
g(t, x+2\pi ) = g(x)\ \Rightarrow\
u(x+2\pi ) = u(x) M\,, \
{\bar u} (x+2\pi ) = {\bar u} (x) {\bar M} 
\ee 
with {\em equal monodromies}, $\bar M = M\,.$ The symplectic 2-form
\be
\lb{sympl2}
\Omega^{(2)} (g, j^0) = \int\limits_{{\Bbb S}^1\; (t={\rm const})}
\omega = {1\over 2}{\rm tr}\int_{-\pi}^{\pi} dx
( ( i j^0 - {k\over{2\pi}} g^{-1}{\partial}_x g ) ( g^{-1} dg )^2 - 
i ( d j^0 ) g^{-1} dg ) 
\ee
is expressed as a sum of two chiral 2-forms involving the monodromy,
\ba
\lb{omega}
&&\Omega^{(2)} (g, j^0)\, =\,
\Omega (u, M) - \Omega({\bar u}, M)\,,\\
&&{\Omega}(u\,,\, M)=
{k\over {4\pi}}\,{\rm tr}\,\{ \int\limits_{-\pi}^\pi dx
\partial \left( u^{-1} du\right) u^{-1} d u - 
u(-\pi )^{-1} du(-\pi ) dM M^{-1} + \rho (M)\}\,.\nonumber
\ea
The 2-form ${\rm tr}\;{\rho}(M)\,$ is
only restricted by the requirement that $\Omega\,$ is closed:
\be
\lb{closed}
d\Omega (u, M) = 0\quad\Leftrightarrow\quad
d\,{\rm tr}\, \rho (M)={1\over 3}\;{\rm tr} \left( dMM^{-1}\right)^3 \,.
\ee
The different choices of $\rho\,$ consistent with (\ref{closed})
correspond to different non-degenerate solutions of the classical
Yang-Baxter equation (YBE). The associated classical $r$-matrices
enter the Poisson bracket (PB) relations for the chiral dynamical
variables \cite{G, FG, FHT1}. A standard choice corresponds to
the Gauss decomposition $M = M_+ M_-^{-1}\,$ of $M\,$ ($M_+\,$
and $M_-^{-1}\,$ involving identical Cartan factors); then
\be
\lb{rho}
\rho (M) = {\rm tr} \left( M_+^{-1} d M_+ M_-^{-1} dM_- \right)\,.
\ee
(For more general choices including a monodromy dependent $r$-matrix see
\cite{BFP}.)

The symplectic form $\Omega^{(2)} (g, j^0)\,$ becomes degenerate when
extended to the space of left and right chiral variables (with a common
monodromy) $\left( u,\; \bar u ,\; M \right)\,.$
This is due to the non-uniqueness of the decomposition (\ref{g}): 
$g(t, x)\,$ does not change under constant right shifts of its chiral
components 
( $u \to u h\,,\ {\bar u} \to {\bar u} h\,,\ M \to
h^{-1} M h\,,\quad h\in G\,$).
We restore non-degeneracy by further extending the phase space introducing 
independent monodromies 
$M\,$ and ${\bar M}\,$ for $u\,$ and ${\bar u}\,$,
respectively, thus completely decoupling the left and right sectors.
The price is that $M\,$ and $\bar M\,$ satisfy Poisson bracket relations
of opposite sign (cf. (\ref{omega})) and monodromy invariance is only
restored in a weak sense -- when $g\,$ is applied to physical states in
the quantum theory.

Quantization is performed by requiring that it respects all
symmetries of the classical chiral theory.
Apart from conformal invariance and invariance under
periodic left shifts the $( u , M )\,$ system
admits a Poisson--Lie symmetry under constant right shifts
\cite{S-T-S, G, FG, AT, FHT1}
which gives rise to a quantum group symmetry in the
quantized theory. This requires passing from the classical to the
quantum $R$-matrix which obeys the quantum YBE: $R_{12} R_{13}
R_{23} = R_{23} R_{13} R_{12}\,$. We end up with quadratic
exchange relations for the chiral variables:
$$
P \, u_1(y)\, u_2(x) = u_1(x)\, u_2(y)\, \R (x-y)\,,\quad
{\bar u}_1(y)\,{\bar u}_2(x) \,P = {\R}^{-1}(x-y)\, {\bar u}_1(x)\,
{\bar u}_2(y)\,,
$$
\ba
\lb{uuR2}
\R (x) =
\left\{
\begin{array}{ccc}
\R & {\rm for} & x > 0 \\
P & {\rm for} & x = 0 \\
{\R}^{-1} & {\rm for} & x < 0
\end{array}
\right .
\ea
(and associated relations for the monodromy $M=u(\pi )^{-1}
u(-\pi )\,$). Here we are using the standard tensor product
notation $u_1 u_2 = u\otimes u\,,\
R_{13} = (R^{\a_1\a_3}_{\b_1\b_3}\d^{\a_2}_{\b_2} )\,$ etc. (see \cite{FRT}),
$P\,$ stands for permutation, so that the first equation
(\ref{uuR2}) is a shorthand for $u^B_\a (y) u^A_\b (x) =
u^A_{{\a}'}(x) u^B_{{\b}'}(y) \R^{{\a}'{\b}'}_{\a\b}(x-y)\,;\
\R\,$ is the braid operator related to the 
standard Drinfeld-Jimbo \cite{Dr, Jimbo} $U_q(sl_n)\ R$-matrix 
by $\R = RP\,$. The quantum YBE for $R\,$ implies the braid relation
\be
\lb{braid}
\R_{12}\R_{23}\R_{12} = \R_{23}\R_{12}\R_{23}\,.
\ee
The exchange relations for $u\,$ are invariant under the left
coaction of the quantum group $SL_q(n)\,$ \cite{HIOPT}. A dual
expression of this property is the $U_q(sl_n)\ (\;\equiv\, U_q )\,$
invariance of (\ref{uuR2}). The $U_q\,$ Chevalley generators $E_i, F_i,
q^{H_i}\,$ can be identified with the elements of the triangular
matrices $M_\pm\,$ \cite{FRT, FHT3}. The $R$-matrix exchange
relations among (elements of) $M_\pm\,$ 
\be
\lb{invcond}
[\R , (M_\pm )_1 (M_\pm )_2 ] = 0 = [\R , ({\bar M}_\pm )_2
({\bar M}_\pm )_1 ]
\ee
imply the CR for the $U_q\,$ generators:
\be
q^{H_i} E_j = E_j q^{H_i+c_{ij}}\,,\ q^{H_i} F_j = F_j q^{H_i-c_{ij}}\,,\
c_{ij} =
\left\{
\begin{array}{ccc}
2 & {\rm for} & i=j \\
-1 & {\rm for} & |i-j|=1 \\
0 & {\rm for} & |i-j|\ge 2
\end{array}
\right .
\,,\nonumber
\ee
\be
\lb{Uq}
[E_i , F_j] =\delta_{ij} [H_i ]\,,\quad X_i X_{i\pm 1} X_i = X_i^{[2]}
X_{i\pm 1} + X_{i\pm 1} X_i^{[2]}\,,\ \  X_i= E_i, F_i
\ee
where
\be
\lb{[m]}
X^{[m]} := {1\over{[m]!}} X^m\quad ([m]! = [m-1]! [m]\,,\
[m]={{q^m -\bq^m}\over{q-\bq}}\,,\ [0]! =
1\,;\ \bq \equiv q^{-1} )\,.
\ee
The $U_q(sl_n)\ \R$-matrix can be written in the form
\be
\lb{R}
q^{1\over n} \R = q\id - A\,,\ \
A^{\a_1 \a_2}_{\b_1 \b_2} =
q^{\epsilon_{\a_2 \a_1}} \delta^{\a_1 \a_2}_{\b_1 \b_2} -
\delta^{\a_1 \a_2}_{\b_2 \b_1}\,, \ \
q^{\epsilon_{\a_2\a_1}} =
\left\{
\begin{array}{ccc}
q & {\rm for} & \a_2 > \a_1 \\
1 & {\rm for} & \a_2 = \a_1 \\
\bq & {\rm for} & \a_2 < \a_1
\end{array}
\right .
\,.
\ee
The {\em $q$-antisymmetrizer} (see, e.g., \cite{Jimbo}) $A\,$ is a
(non-normalized) projector,
\be
\lb{TL}
A^2\; =\; [2]\, A\quad\quad (\; [2] = q + \bq \; )
\ee
satisfying, due to the braid relation (\ref{braid}),
\be
\lb{AAA-A}
A_{12}\; A_{23}\; A_{12}\; -\; A_{12}\; =\;
A_{23}\; A_{12}\; A_{23}\; -\; A_{23}\,.
\ee
Eq. (\ref{TL}) is equivalent to the {\em Hecke property}
\be
\lb{Hecke}
q^{2\over n}\R^2=\id + (q-\bq )\; q^{1\over n} \R\,
\ee
that does not hold for
a deformation of a simple Lie group different from $SL(n)\,$.
The parameter $q\,$ is expressed in terms of the height
$h=k+n\ ( \ge n\,$) by
\be
\lb{q}
q=e^{i{{\pi}\over h}}\quad (q^{1\over n} = e^{i { {\pi}\over{nh} } } )
\quad\Rightarrow\quad q^h = -1\,;
\ee
here $k \in {\Bbb Z}_+\,$ is the quantized coupling constant of
(\ref{can3}), (\ref{eqmo}) identified with the Kac-Moody level.

\section{Chiral vertex operators and zero modes. The quantum
matrix algebra}
\setcounter{equation}{0}
\renewcommand{\theequation}{\thesection.\arabic{equation}}

Let $\{ v^{(i)}, \, i=1,\ldots , n\}$ be a symmetric {\em
"barycentric basis"} of (linearly dependent) real traceless
diagonal matrices :
\be
\lb{bary}
(v^{(i)})^j_k = \left(\d_{ij} - {1\over n}\right)\d^j_k\quad\Rightarrow\quad
\sum_{i=1}^n\, v^{(i)} = 0\,.
\ee
The simple $sl(n)$ (co)roots ($\a_i^{\vee} =$) $\a_i$ and the
corresponding (co)weights $\Lambda^{(j)}$ are expressed in terms
of $v^{(i)}$ as follows:
\be
\lb{roots}
\a_i=v^{(i)}-v^{(i+1)}\,,\quad \Lambda^{(j)} = \sum_{i=1}^j v^{(i)}
\qquad (\, \Rightarrow (\Lambda^{(j)} | \a_i ) = \d^j_i )
\ee
(the inner product of two matrices coinciding with the trace of
their product).
A {\em shifted dominant weight}
\be
\lb{weight}
p=\Lambda +\rho\,,\ \
\Lambda = \sum_{i=1}^{n-1}\,\lambda_i\,\Lambda^{(i)}\,,\ \
\lambda_i\in {\Z}_+\,,\ \
\rho = \sum_{i=1}^{n-1}\,\Lambda^{(i)}\
( = {1\over 2}\sum_{\a >0}\,\a )
\ee
can be conveniently parametrized by $n\,$ numbers $\{ p_i \}\,$
($\;\sum_{i=1}^n\; p_i = 0\;$) satisfying 
\be
\lb{dominant}
p_{i\, i+1} = \lambda_i +1 \in {\Bbb N}\,,\ \
i=1,2,\ldots ,n-1\quad {\rm where}\quad p_{ij} \equiv p_i - p_j\,.
\ee
(The non-negative integers $\lambda_i = p_{i\, i+1} - 1\,$ count
the number of columns of length $i$ in the Young tableau that
corresponds to the IR of highest weight $p$ of $SU(n)$ -- see,
e.g., \cite{Ful}.)
Dominant weights $p$ also label
highest weight representations of $U_q\,.$ For integer heights
$h\, (\ge n )\,$ and $q$ satisfying (\ref{q}) these are (unitary)
irreducible if $( n-1\le\, )\, p_{1n}\, \le h\,$. The {\em quantum
dimension} of such an IR is given by \cite{CP}
\be
\lb{qdim}
d_q(p) = \prod_{i=1}^{n-1} \{\, {1\over{[i]!}}\,\prod_{j=i+1}^n\,
[p_{ij}]\}\quad (\,\ge 0\quad{\rm for}\quad
p_{1n}\,\le h )\,.
\ee
For $q\rightarrow 1\ (h\rightarrow \infty )\,, \ [m]\rightarrow m\,$
we recover the usual (integral) dimension of the IR under consideration.

Energy positivity implies that the state space of the chiral
quantum WZNW theory is a direct sum of (height $h\,$) ground state modules
${\cal H}_p\,$ of the Kac-Moody algebra
${\widehat{su}}(n)$ with a finite multiplicity:
\be
\lb{space}
{\cal H}\, =\, \bigoplus_p\,{\cal H}_p\otimes {\cal V}_p\,,\quad
{\rm dim}\,{\cal V}_p < \infty\,.
\ee
Each "internal space" ${\cal V}_p\,$ carries a representation
of $U_q\,$ and is
an eigensubspace of all $U_q\,$ Casimir invariants (whose eigenvalues
are polynomials in $q^{p_i})\,.$

The analysis of the axiomatic construction of the quantum field theory
generated by a chiral current algebra \cite{KZ, BPZ, GW, TK, STH} yields
the following properties of the representations of $\;U_q\,$ 
in the space (\ref{space}):

(i) the ideal generated by $E_i^h\,,\ F_i^h\,,\ [h H_i]\,$ is
represented trivially;

(ii) the {\em integrable highest weight representations}
corresponding to positive integer $p_{i\; i+1}\,$ and $ n-1\le
p_{1n} < h\,$ appear in pair with representations of weight ${\tilde p}\,$
where 
\be
\lb{tild}
{\tilde p}_{12} = 2h - p_{12}\ \ 
{\rm for}\ n=2\,;\quad 
{\tilde p}_{12} = h - p_{23}\,,\ 
{\tilde p}_{23} = h - p_{12}\ \ 
{\rm for}\ n=3\,,\ {\rm etc.} 
\ee
which corresponds to the highest weight of a subspace of singular
vectors in the Verma module ${\cal H}_p\,$.

Each ${\cal H}_p\,$ in the direct sum (\ref{space}) is a graded
vector space,
\be
\lb{grad}
{\cal H}_p \, =\, \oplus_{\nu =0}^\infty\; {\cal H}_p^\nu\,,\quad
\left( L_0 - \Delta_h (p) - \nu\right)\; {\cal H}_p^\nu\, =\, 0\,,
\quad {\rm dim}\, {\cal H}_p^\nu < \infty\,,
\ee
where $L_0\,$ is the chiral (Virasoro) energy operator. Here
${\cal H}_p^0\,$ spans
an IR of $su(n)\,$ of (shifted) highest weight $p\,$.
The {\em conformal dimension}
(or {\em conformal weight}) $\Delta_h (p)\,$ is proportional to the
($su(n)$-) second order Casimir operator $|p|^2 - |\rho |^2\,,$
and $\Delta_h ({\tilde p}) - \Delta_h (p)\,$ is an integer; the
vacuum weight is zero:
\ba
&&\lb{Delta}
2h\Delta_h(p) = |p|^2 - |\rho |^2 =
{1\over n} \sum_{1\le i < j\le n}\,p_{ij}^2 - {{n(n^2-1)}\over
{12}}\,,\nonumber\\
&&\Delta_h ({\tilde p}) - \Delta_h (p) = h-p_{1n}\ \ 
{\rm for}\ n=2,3;\  \quad \Delta_h
(p^{(0)}) = 0 \ \ {\rm for}\ \ p^{(0)}_{i\;i+1} = 1 \,.
\ea
The eigenvalues of the braid operator $\R\,$ are expressed as
exponents of differences of conformal dimensions \cite{FHT2,
FHT3, HIOPT}. This yields (\ref{q}) for $q\,$.

We shall split the $SU(n)\times SL_q(n)\,$ covariant field $u(x) =
\left( u^A_\a (x)\right)\,$ into factors which intertwine
separately different ${\cal H}_p\,$ and ${\cal V}_p\,$ spaces.
A CVO $u_j(x,p)\,$
is defined as an intertwining map between ${\cal H}_p\,$ and
${\cal H}_{p+v^{(j)}}\,$ (for each $p$ in the sum (\ref{space})).
Noting that ${\cal H}_p\,$ is an eigenspace of $e^{2\pi i\, L_0}\,$,
\be
\lb{specL}
{\rm Spec}\, L_0\, |_{{\cal H}_p}\,\subset \Delta_h(p)\, +{\Z}_+ \quad
\Rightarrow\quad \{ {e}^{2\pi i\, L_0}\, - \,
{e}^{2\pi i\,\Delta_h(p)}\,\}\, {\cal H}_p\, = \, 0\,,
\ee
we deduce that $u_j(x,p )\,$ is an eigenvector of the monodromy
automorphism,
\be
\lb{monodromyauto}
u_j(x+2\pi ,p)=e^{-2\pi iL_0}\, u_j(x,p)\,e^{2\pi iL_0}\,=
u_j(x,p)\,\mu_j(p)
\ee
where,
in view of (\ref{specL}), we find
\be
\lb{mu}
\mu_j(p) \,:=\, e^{2\pi i \{\Delta_h(p) - \Delta_h(p+v^{(j)})\}}\,=\,
q^{{1\over n}-1-2p_j}\,.
\ee
The monodromy matrix is diagonalizable whenever its
eigenvalues (\ref{mu}) are all different. The exceptional points are
those $p$ for which there exists a pair of indices $1\le i <j\le n\,$
such that $q^{2\,p_{i j}}\, =\, 1\,$, since we have
\be
\lb{exceptional}
{{\mu_j(p)}\over{\mu_i(p)}}\,=\,q^{2\,p_{i j}}
\,.
\ee
According to (\ref{qdim}) all such "exceptional" ${\cal V}_p\,$
have zero quantum dimension. In particular, for the "physical
IR", characterized by $p_{1n}\, <\,h\,,\ M\,$ is diagonalizable.

The {\em quantum matrix} $a\,=\, \left(a^j_\a\,,\ j,\a=1,\ldots
,n\right)\,$ is defined to relate the $SL_q(n)\,$ covariant field
$u(x)\,=\,\left( u^A_\a (x)\right)\,$ with the CVO
$u^A_j(x,p)\,$ realizing the so called
vertex--IRF (interaction-round-a-face) transformation \cite{BBB}
\be
\lb{vIRF}
u(x)\,=\,a^j\,u_j(x,p)\quad \left(\,u_j\,
=\,(u^A_j)\,,\,a^j\,=\,(a^j_\a)\,\right)\,.
\ee
The zero mode operators $a^j\,$ are defined to intertwine the
finite dimensional $U_q\,$ modules ${\cal V}_p\,$ of
(\ref{space}):
\be
\lb{aV-a}
a^j\,:\ {\cal V}_p\,\rightarrow\, {\cal V}_{p+v^{(j)}}\,.
\ee
For $n=2\,$ the operators $a^2\,$ were treated as annihilation operators
\cite{FHT2}. In fact, in the case of irreducible ${\cal V}_p\,$ realized
(for any $n\ge 2\;$) in
the Fock space representation of the quantum matrix algebra introduced in
Section 3 below, we have the annihilation property 
\be
\lb{aV-b}
a^j\; {\cal V}_p = 0 \quad {\rm if}\quad p_{j-1} = p_j+1\,,\quad j > 1 
\ee
(thus $a^j\,{\cal V}_p\, $ is zero unless $p+v^{(j)}\,$ is again
a dominant weight).

The zero modes $a^j_\a\,$ commute with the currents,
thus leaving the $\widehat{su}(n)\,$
modules ${\cal H}_p$ unaltered. The order of factors in
(\ref{vIRF}) is dictated by the requirement that the  
the commuting {\em operators} $p_i\;,\  i=1,\dots ,n\,,$ 
the components of the argument $p\,$ of $u_j\;,$
are proportional to the unit operators on ${\cal H}_p\,$ with 
eigenvalues corresponding to the label of the module. 
We note that
\be
\lb{qa}
q^{p_i}\,a^j\,=\,a^j\,q^{p_i+\d_{ij}-{1\over n}}\quad\Rightarrow\quad
a^j\,u_i(x,p)\,=\,u_i(x,p-v^{(j)})\, a^j\,.
\ee
The $U_q\,$ covariance properties of $a^i_\a\,$ can be read off
their exchange relations with the Gauss components $M_\pm\,$ of
$M\,$ \cite{FHT3, FHIPT}:
\ba
&&[E_a,\,a^i_\a]\,=\,\d_{a\,\a-1}\,a^i_{\a-1}\,q^{H_a}\,,\
a=1,\ldots ,n-1 \,,\nonumber\\
&&F_a\,a^i_\a\,-q^{\d_{a\,\a-1}-\d_{a\,\a}}\,a^i_\a\,F_a\,=
\,\d_{a\,\a}\,a^i_{\a+1}\,;\\
&&q^{H_a}\,a^i_\a\,=\,a^i_\a\,q^{H_a+\d_{a\,\a}-\d_{a\,\a-1}}\,.\nonumber
\ea
Comparing (\ref{per}), (\ref{monodromyauto}) and (\ref{vIRF}) we deduce
that
the zero mode matrix $a\,$ diago-nalizes the monodromy (whenever
the quantum dimension (\ref{qdim}) does not vanish); setting
$a\,M\,=\,M_p\,a\,$
we find (from the above analysis of Eqs. (\ref{monodromyauto}) -
(\ref{exceptional})) the implication
\be
d_q(p)\,\ne\,0\quad\Rightarrow\quad
\left(M_p\right)^i_j\,=\,\d^i_j\,
\mu_j(p - v^{(j)})\,, \quad
\mu_j(p - v^{(j)}) = q^{1-{1\over n}-2{p}_j}\,.
\ee
(A careful study of the case of vanishing $d_q(p)\,$ and
non-diagonalizable $M\,$ is still lacking.)

The exchange relations (\ref{uuR2}) for $u\,$ given by (\ref{vIRF}) can be
translated into quadratic exchange relations for the "$U_q\,$
vertex operators" $a^i_\a\,$ provided we assume standard braid relations
for the CVO $u(x,p )\,$:
\be
\lb{uuRp}
u^B_i(y,p -v^{(i)})\,
u^A_j(x,p -v^{(i)}-v^{(j)})\, =\,
u^A_k(x,p -v^{(k)})\,
u^B_l(y,p -v^{(k)}-v^{(l)}) \, \R (p)^{kl}_{ij}
\ee
(what is important here is that $\Rp\,$ depends only on $p\,$). 
An analysis of chiral 4-point blocks \cite{FHIPT} shows that
Eq.(\ref{uuRp}) is indeed satisfied.

A straightforward computation then gives
\be
\lb{Rpaa=aaR}
\R (p )\, a_1\, a_2\, =\, a_1\, a_2\, \R\,.
\ee
Associativity of triple tensor products of quantum matrices together with
Eq.(\ref{braid}) for $\R\,$
yield, as a consistency condition of (\ref{Rpaa=aaR}), the
quantum dynamical YBE for $\Rp\,$ (first studied in \cite{GN}):
\be
\lb{QDYBE}
\DR{12}\, \R_{23}(p-v_1)\, \DR{12}\, =\,
\R_{23}(p-v_1)\, \DR{12}\, \R_{23}(p-v_1)
\ee
where we use again the succint notation of Faddeev et al. \cite{FRT}:
\be
\lb{R23}
\left( \R_{23}(p-v_1)\right)^{i_1 i_2 i_3}_{j_1 j_2 j_3} \, =\,
\d^{i_1}_{j_1}\,\R (p-v^{(i_1)})^{i_2 i_3}_{j_2 j_3}\,.
\ee
In deriving (\ref{QDYBE}) from (\ref{Rpaa=aaR}) we use (\ref{qa}). 
(The quantum dynamical YBE (\ref{QDYBE}) is only {\em
sufficient} for the consistency of the quadratic matrix algebra relation
(\ref{Rpaa=aaR}); it would be also necessary if the matrix $a\,$ were
invertible -- i.e., if $d_q(p)\ne 0\,$.) 

The property of the operators $\R_{i\, i+1}(p)\,$ to generate a
representation
of the braid group is ensured by the additional requirement (reflecting
the commutativity of the braid group generators $B_i\,$ and
$B_j\,$ for $|i-j|\ge 2\,$)
\be
\lb{Rpvv}
\R_{12} (p+v_1+v_2 )\, =\, \R_{12}(p)\quad \Leftrightarrow\quad
\R^{ij}_{kl} (p) a^k_\a a^l_\b \, =\,
a^k_\a a^l_\b \,\R^{ij}_{kl} (p) \,.
\ee
The Hecke algebra condition (\ref{Hecke})  
follows from the analysis of braiding properties of
conformal blocks \cite{FHIPT}.

We shall adopt here the following solution of Eq.(\ref{QDYBE})
and the Hecke algebra condition
(presented in a form similar to (\ref{R})):
\be
\lb{A1}
q^{1\over n}\,\Rp\, =\, q\id - A(p)\,,\quad
A(p)^{ij}_{kl}\, =\, {{[p_{ij}-1]}\over{[p_{ij}]}}\, \left(
\d^i_k \d^j_l - \d^i_l\d^j_k\right)\,.
\ee
$A(p)\,$ satisfies similar
relations as $A\,$ (cf. (\ref{TL}), (\ref{AAA-A}));
in particular,
\be
\lb{A2}
[p_{ij}-1]\, +\, [p_{ij}+1]\, =\, [2]\,[p_{ij}]\quad\Rightarrow\quad
A^2 (p) \, =\, [2]\, A(p)\,.
\ee
According to \cite{HIOPT} the general
$SL(n)$-type dynamical $R$-matrix \cite{I}
can be obtained from (\ref{A1}) by either a dynamical analog of
Drinfeld's twist
\cite{D2} (see Lemmas 3.1 and 3.2 of \cite{HIOPT}) or by a canonical
transformation
$p_i \to p_i +c_i\,$ where $c_i\,$ are constants (numbers) such that
$\sum_{i=1}^n c_i = 0\,$. The interpretation of the eigenvalues of
$p_i\,$ as (shifted) weights (of the corresponding representations of
$U_q\,$) allows to dispose of the second freedom.

Inserting (\ref{A1}) into the exchange relations (\ref{Rpaa=aaR}) allows
to present the latter in the following explicit form:
\be
\lb{aa1}
[a^i_\a , a^j_\a ] = 0\,,\quad
a^i_\a a^i_\b = q^{{\epsilon}_{\a\b}} a^i_\b a^i_\a
\ee
\be
\lb{aa2}
[p_{ij}-1]\, a^j_\a a^i_\b = [p_{ij}] a^i_\b a^j_\a -
q^{{\epsilon}_{\b\a}p_{ij}} a^i_\a a^j_\b\ \ {\rm for}\ \ \a\ne\b\ \
{\rm and}\ i\ne j,
\ee
where $q^{\epsilon_{\b\a}}\,$ is defined in (\ref{R}).
There is, finally, a relation of order $n$ for $a^i_\a\,$, derived from
the following basic property of the {\em quantum determinant}:
\be
\lb{qdet}
\det (a)\, =
{1\over{[n]!}}\, {\varepsilon}_{\scriptscriptstyle i_1 \dots i_n}
a^{i_1}_{\a_1} \dots a^{i_n}_{\a_n}\,
{\cal E}^{\scriptscriptstyle \a_1 \dots \a_n}
\ee
where
\be
\lb{qge}
{\cal E}^{\scriptscriptstyle \a_1 \dots \a_n}\
(= {\cal E}_{\scriptscriptstyle \a_1 \dots \a_n} ) =
\bq^{{n(n-1)}\over 4} (-q)^{\ell (\sigma )}\quad {\rm for}\quad
\sigma = {n_{~}, \dots ,1_{~}\choose \a_1, \dots ,\a_n}\,,
\ee
$\ell(\sigma)$ being the length of the permutation
$\sigma\,; {\varepsilon}_{\scriptscriptstyle i_1 \dots i_n} \,$ is the
undeformed Levi-Civit\`a tensor normalized by
${\varepsilon}_{\scriptscriptstyle n\dots 1} = 1 \,$; the ratio
$\det (a) \left( \prod_{i<j} [p_{ij}] \right)^{-1}\,$ belongs to the
centre of the {\em quantum matrix algebra}
${\cal A} = {\cal A} (\Rp , \R )\,$ -- the associative algebra with generators
$q^{p_i}\,,\ a^j_\alpha\,$ and relations $q^{p_i} q^{p_j} = q^{p_j}
q^{p_i}\,,\ \,
\prod\limits_{i=1}^n q^{p_i} = 1,\,$ as
well as (\ref{qa}) and (\ref{Rpaa=aaR}) (see Corollary 5.1 of Proposition 5.2 of
\cite{HIOPT}).
It is, therefore, legitimate to normalize the quantum determinant setting
\be
\lb{qdnorm}
\det (a)\, =\,
\prod_{i<j}\, [p_{ij}]
\equiv \, {\cal D} (p)\,.
\ee
Clearly, it is proportional (with a $p$-independent  
positive factor) to the quantum dimension (\ref{qdim}).

We shall use in what follows the intertwining properties of the product
$a_1\dots a_n\,$ (see Proposition 5.1 of \cite{HIOPT}):
\ba
\lb{int}
{\varepsilon}_{\scriptscriptstyle i_1 \dots i_n} \,
a^{i_1}_{\a_1}\dots
a^{i_n}_{\a_n}\, &=& \, {\cal D}(p)\,
{\cal E}_{\scriptscriptstyle \a_1 \dots \a_n}\,,\\
a^{i_1}_{\a_1}\dots a^{i_n}_{\a_n}\,
{\cal E}^{\scriptscriptstyle \a_1 \dots \a_n}\,
&=&\, {\varepsilon}^{\scriptscriptstyle i_1 \dots i_n} (p) \,
{\cal D}(p)\,.
\ea
Here ${\varepsilon}^{\scriptscriptstyle i_1 \dots i_n} (p)\,$ is the
dynamical Levi-Civit\`a tensor given by
\be
\lb{dLC}
{\varepsilon}^{\scriptscriptstyle i_1 \dots i_n}(p)\,
=\, (-1)^{\ell(\sigma )}\,\prod_{1\le\mu <\nu\le n}
{ {[p_{i_\mu i_\nu} -1]}\over{[p_{i_\mu i_\nu}]} }\,.
\ee

We note that the "monodromy subalgebra" of ${\cal A}\,,$ i.e., the commutant of
$\{ q^{p_i}\,,\ i=1,\dots ,n\;\}\,,$ is generated by $U_q\,$ and $\{ q^{p_i}\;\}\,.$  

\section{Fock space representation of ${\cal A}\,$ and its
finite dimensional quotient space}
\setcounter{equation}{0}
\renewcommand{\theequation}{\thesection.\arabic{equation}}

The {\em Fock space} ${\cal F} = {\cal F} ({\cal A})\,$
for the quantum matrix algebra ${\cal A}\,$
is defined as a (reducible) $U_q\,$ module with a 1-dimensional
$U_q$-invariant subspace $\{{\Bbb C}\, |0{\cal i} \}\,$ where the
vacuum vector $|0{\cal i} \equiv |{p^{(0)}}, 0{\cal i}\,$ is cyclic for
${\cal A}\,$ and is annihilated by all monomials in $a^i_\a\,$ which
do not correspond to a Young tableau:
${\cal A}\, |0{\cal i} = {\cal F}\,,$ and
\ba
\lb{Y}
&&{\cal P}_{m_n} (a^n_\a )\dots {\cal P}_{m_1} (a^1_\a ) |0{\cal i}
= 0\\
&&{\rm for} \quad {\cal P}_{m_i} (a^i_\a )= (a^i_1)^{m_{i1}}\dots
(a^i_n)^{m_{in}}\quad
{\rm unless}\quad m_1\ge m_2\ge \dots\ge m_n\nonumber
\ea
where $m_{i s}\in {\Bbb Z}_+\,$ and $m_i = m_{i1}+\dots +m_{in}\,.$
The $U_q\,$ structure of ${\cal F}\,$ is given by the following
statements \cite{FHIPT}.
\vspace{3mm}

\noindent
{\bf Proposition 3.1}~{\em The Fock space ${\cal F}\,$ is a
direct sum of finite dimensional $U_q\,$ modules ${\cal F} (p)\,$
spanned by monomials of the type
\be
\lb{span}
{\cal P}_{\lambda_{n-1}} (a^{n-1}_\a )
{\cal P}_{\lambda_{n-2}+\lambda_{n-1}} (a^{n-2}_\a )
\dots
{\cal P}_{\lambda_1+\dots +\lambda_{n-1}} (a^1_\a )
|0{\cal i}\ (\in {\cal F}(p) )
\ee
where $\lambda_i = p_{i\;i+1}-1 \in {\Bbb Z}_+\,$ and the
factors ${\cal P}_m(a^i_\a )\,$ are defined in (\ref{Y}). For
$p_{1n}\le h\,$ the resulting $U_q\,$ modules are irreducible.}

\vspace{3mm}

\noindent
{\bf Theorem 3.2}~{\em The space ${\cal F}\,$ admits a
(symmetric) $U_q\,$ invariant bilinear form
${\cal h}~,~{\cal i}\,$ determined by the conditions
\be
\lb{bil}
{\cal h}\Phi , X \Psi {\cal i} =
{\cal h} X' \Phi , \Psi{\cal i}\,,\ \forall X\in {\cal A}\,,\quad
{\cal h} 0 | 0{\cal i} = 1
\ee
where the "transposition" $X \to X'\,$ is a linear antiinvolution
on ${\cal A}\,$ such that
\be
\lb{transp1}
E_i' = F_i\; q^{H_i -1}\,,\quad F_i' = q^{1-H_i} E_i\,,\quad
(q^{H_i})' = q^{H_i}\,,\quad (q^{p_{ij}})' = q^{p_{ij}}
\ee
and
\ba
\lb{transp2}
(a^i_\a )' = 
{\tilde a}_i^\a 
/{\cal D}_i(p)\,;\quad
{\tilde a}_i^\a &=& {1\over{[n-1]!}}
{\cal E}^{\scriptscriptstyle \a\a_1 \dots \a_{n-1}}
{\varepsilon}_{\scriptscriptstyle i i_1 \dots i_{n-1}}
a^{i_1}_{\a_1}\dots a^{i_{n-1}}_{\a_{n-1}}\,,\nonumber\\
{\cal D}_i(p) &=& \prod\limits_{{j<\ell}\atop{j\ne i\ne\ell}}
[p_{j\ell} ]\,.
\ea
For $p_{i\;i+1}\le h\,$ the $U_q\,$ modules ${\cal F}(p)\,$ are
unitary with respect to a sesquilinear inner product $(~,~)\,$
such that
\be
\lb{inner}
(\Phi , X \Psi ) = (X^* \Phi , \Psi )\quad {\it for}\quad
X \in U_q\,,\quad (0 | 0) = 1
\ee
where the hermitean conjugation $X \to X^*\,$ is an antilinear
involutive antihomomorphism of $U_q\,$ satisfying
\be
\lb{*}
E_i^* = F_i\,,\quad F^*_i = E_i\,,\quad (q^{H_i})^* =
\bq^{H_i}\,,\quad (q^{p_{ij}})^* = \bq^{p_{ij}}\,.
\ee
The bilinear form ${\cal h}~,~{\cal i}\,$ is majorized by the
scalar product $(~,~)\,$:
\be
\lb{maj}
|{\cal h}\Phi,\Psi{\cal i}|^2 \le (\Phi ,\Phi )(\Psi ,\Psi )\quad
{\it for}\quad \Phi , \Psi \in {\cal H}' :=
\oplus_{p_{i\;i+1}\le h} {\cal F}(p)\,.
\ee
There exists a basis $\{\Phi_\nu \}\,$ 
(e.g., the canonical basis described below, for $n=2,3\,$) in each
${\cal F}(p)\subset {\cal H}'\,$ of eigenvectors of $q^{H_i}\,$
such that $|{\cal h}\Phi_\nu,\Phi_\nu{\cal i}| = 
(\Phi_\nu ,\Phi_\nu )\,$.}

\vspace{3mm}

We note that while the transposition is a coalgebra homomorphism,
-- i.e., it preserves the coproduct $\Delta\,$,
\be
\lb{Dt}
\Delta (X)' = \Delta (X')\quad ({\rm where}\ \
\Delta ({M_\pm}^{\a~}_{~\b}) =
{M_\pm}^{\a~}_{~\sigma}\otimes{M_\pm}^{\sigma~}_{~\b} )\,,
\ee
the conjugation reverses the order of factors in tensor products.

The algebra ${\cal A}\,$ admits, for $q\,$ given by (\ref{q}), a
large ideal
\be
\lb{ideal}
{\cal I}_h \ ( = {\cal A}\, {\cal I}_h\, {\cal A} \subset
{\cal A} )\, \quad {\rm generated\  by}\ \
(a^i_a )^h\,\ 
{\rm and}\ [h p_{ij}]\,.
\ee
This is a consequence (for $m=h\,$) of the general exchange relation
\be
\lb{genex}
[p_{ij}-1](a^j_\a )^m a^i_\b = [p_{ij}+m-1] a^i_\b
(a^j_\a)^m - q^{\epsilon_{\b\a} (p_{ij}+m-1)}
[m](a^j_\a)^{m-1} a^i_\a a^j_\b
\ee
which follows from (\ref{aa2}), (\ref{aa1}).
Both forms (\ref{bil}) and
(\ref{inner}) are degenerate, their kernel containing ${\cal I}_h
{\cal F}\,$:
\be
\lb{degen}
{\cal h}\Phi , {\cal I}_h \Psi{\cal i} = 0 =
(\Phi , {\cal I}_h \Psi )\quad \forall\;\Phi ,\,\Psi\,\in {\cal
F}\,.
\ee
The quotient space ${\cal F}_h:= {\cal F} /{\cal I}_h {\cal F}\,$
is finite dimensional: it is isomorphic to the subspace of ${\cal
F}\,$ spanned by vectors of the form (\ref{span}) with $\lambda_1
+\dots +\lambda_{n-1}\,\le n(h-1)\,$, -- i.e., $p_{1n} < nh\,$.
It carries a representation of the factor algebra ${\cal A}_h\,$
whose $p$-invariant subalgebra (of elements $X\,$ satisfying
$q^{p_i} X\;\bq^{p_i} = X\,$) is the
"finite dimensional (or restricted) quantum group" $U_h\,$:
\be
\lb{restr}
{\cal A}_h = {\cal A}/{\cal I}_h \;\supset\; U_h = U_q /
{\cal I}_h^U\,,\quad {\cal I}_h^U = {\cal I}_h\cap U_q\,,
\ee
${\cal I}_h^U\,$ being generated by
$ E_i^h \,,\ F_i^h \,$ and $[h H_i ]\,.$

As an example, and in order to prepare the ground for the
discussion in Section 4, we shall display the canonical basis in
${\cal F}_h\,$ for $n=2\,$. Setting in this case $p_{12}=p\,$ we
can write
\be
\lb{basis2}
|p,m{\cal i}:= (a^1_1 )^m (a^1_2 )^{p-1-m} | 0{\cal i}
\ee
with $m\,$ belonging to the intervals $0\le m<p\,,$ for $0<p\le
h\,,$ and $p-h \le m<h\,$ for $h<p<2h\,,$ respectively. The
$U_q (sl_2 )\,$ properties of this basis are summed up by
\be
\lb{Uqprop2}
(q^{H-2m+p-1}-1)|p,m{\cal i}=0\,,\
E|p,m{\cal i} = [p-m-1]|p,m+1{\cal i}\,,\
F|p,m{\cal i} = [m] |p,m-1{\cal i}\,.
\ee
${\cal F}_h\,$ can be written as a sum of irreducible $U_q (sl_2 )\,$
modules and its dimension can be computed as follows:
\ba
\lb{dimF}
&&{\cal F}_h = \left({\oplus}_{p=1}^h\; {\cal F}(p) \right)\
{\oplus}\ \left( {\oplus}_{p=h+1}^{2h-1} {\tilde{\cal F}}(p) \right)\ \quad
\left(\ {\tilde{\cal F}}(p) = 
{\cal F}({\tilde p}) / {\cal I}_h^U\;{\cal F}({\tilde p})\ \right),\\
&&{\rm dim}\,{\cal F}(p) = p\,,\quad
  {\rm dim}\,{\tilde{\cal F}}(p) = 2h-p\,,\quad
  {\rm dim}\,{\cal F}_h = h^2\,.\nonumber
\ea
Eq.(\ref{transp2}) reduces in this case to
\be
\lb{tr-2}
(a^i_\a )' = {\tilde a}^\a_i = {\cal E}^{\a\b}\varepsilon_{ij}
a^j_\b\,,\ {\rm - i.e.,}\
(a^1_1 )'={\tilde a}^1_1 = q^{1\over 2} a^2_2\,,\
(a^1_2 )'={\tilde a}^2_1 = - \bq^{1\over 2} a^2_1\,,\ {\rm etc.}
\ee
In order to compute the bilinear form (\ref{bil}) in the basis
(\ref{basis2}) one needs the relations
\be
\lb{atilde}
{\tilde a}^1_1 |p,m{\cal i} = q^{m-p+1}[m]|p-1,m-1{\cal i}\,,\quad
{\tilde a}^2_1 |p,m{\cal i} = [p-m-1]|p-1,m{\cal i}\,.
\ee
We deduce that the basis (\ref{basis2}) is orthogonal with respect
to both inner products and verifies the last statement of Theorem
3.2:
\be
\lb{verify}
{\cal h} p,m | p',m'{\cal i} = \delta_{pp'}\delta_{mm'}\,
\bq^{m(p-m-1)}\; (p,m |p,m )\,,\quad (p,m |p,m )=[m]![p-m-1]!\,.
\ee
\vspace{3mm}

\noindent
{\bf Remark 3.1}~The hermitean conjugation (\ref{*}) can be
extended to the entire algebra ${\cal A}\,$
(and to ${\cal A}_h\,$) setting
\be
\lb{a*}
(a^1_1 )^*=q^{{p-H-1}\over 2}{\tilde a}^1_1\ (\equiv a_1\
\Rightarrow\ a^1_1 = a^*_1\,)\,,\
(a^1_2 )^*=q^{{1-p-H}\over 2}{\tilde a}^2_1\ (\equiv a_2\
\Rightarrow\ a^1_2 = a^*_2\,)\,.
\ee

We then verify that products of the type $X X^*\,,$
\ba
\lb{XX*}
a^1_1 (a^1_1 )^* \, (\; = a_1^* a_1 ) = \left[ {{H+p-1}\over
2}\right]\,,\quad
(a^1_1 )^* a^1_1 \, (\; = a_1 a_1^* ) = \left[ {{H+p+1}\over
2}\right]\,,\nonumber\\ \\
a^1_2 (a^1_2 )^* \, (\; = a_2^* a_2 ) = \left[ {{p-H-1}\over
2}\right]\,,\quad
(a^1_2 )^* a^1_2 \, (\; = a_2 a_2^* ) = \left[ {{p-H+1}\over
2}\right]\,,\nonumber
\ea
are positive semidefinite, $X X^* \ge 0\,,$ in ${\cal F}_h\,.$ It
follows that the majorization property (\ref{maj}) extends to the
entire quotient space ${\cal F}_h\,$.
\vspace{3mm}

\noindent
{\bf Remark 3.2}~The canonical basis $\{ |p,m{\cal i}\,,\ 0\le
m\le p-1\;\}\,$ in ${\cal F} (p)\,$ for $p\le h\,$ can be also
expressed in terms of the lowest and highest weight vectors in
${\cal F}(p)\,$
\be
\lb{lhwv}
\left[{{p-1}\atop m}\right] |p,m{\cal i} = E^{[m]}|p,0{\cal i} =
F^{[p-1-m]}|p,p-1{\cal i}\,,
\ee
where we are using the notation $X^{[m]}\,$ of (\ref{[m]}). These
expressions are readily generalized to the case of $U_q(sl_3 )\,$
in terms of Lusztig's canonical basis \cite{L} of (say, 
raising) operators 
\be
\lb{Lus}
E_1^{[m]}E_2^{[\ell ]}E_1^{[k]}\,,\quad E_2^{[k]}E_1^{[\ell
]}E_2^{[m]}\quad{\rm for}\quad \ell\ge k+m
\ee
where the Serre relations imply
\be
\lb{k+m}
E_1^{[m]}E_2^{[k+m]}E_1^{[k]}\, =\,
E_2^{[k]}E_1^{[k+m]}E_2^{[m]}\,.
\ee

\section{The physical zero mode space: a generalized
BRS construction}
\setcounter{equation}{0}
\renewcommand{\theequation}{\thesection.\arabic{equation}}
The zero mode part of the monodromy extended $2D\,$
WZNW model involves the tensor product ${\cal A}_h\otimes{\overline{\cal
A}}_h\,$ of two isomorphic copies of the chiral factor algebra
(\ref{restr}) in the finite dimensional state space
\be
\lb{findimstate}
{\cal H}\, =\, {\cal F}_h \otimes {\overline{\cal F}}_h\,.
\ee
The physical space ${\frak H}\,=\,{\frak H}_{h,n}\,,$ on the other
hand, is known (from the axiomatic treatment of the model) to have
the following properties:

\noindent (i) it is the complexification of the real Hilbert space
${\frak H}_{\Bbb R}\,$ of dimension ${h-1}\choose{n-1}\,$ with
inner product coinciding with the bilinear form ${\cal h}~,~{\cal
i}\,;$

\noindent (ii) it is monodromy invariant \cite{FHT2, FHT3},
\be
\lb{moninv}
a^i_\a\;\{ (M{\bar M}^{-1} )^\a_\b - \delta^\a_\b \}\;
{\bar a}^\b_j \,{\frak H}\, =\, 0\,;
\ee

\noindent (iii) it is invariant under the diagonal (coproduct)
action of $U_q(sl_2)$
\be
\lb{Deltainv}
\{\Delta (X) -\epsilon (X)\}\,{\frak H}\,=\,0\quad{\rm
or,\ explicitly,}\quad\{ ({\bar M}_\pm^{-1} M_\pm )^\a_\b -
\delta^\a_\b \}\,{\frak H}\,=\,0\,,
\ee
where $M_\pm\,$ are the Gauss components of $M\,$ (see Section 1)
defined, in the quantum case, by $q^{n-{1\over n}} M = M_+
M_-^{-1}\,$.

The following question arises: can we construct ${\frak H}\,$ as a
subquotient of ${\cal H}\,$ using a kind of a BRS formalism? A
constructive answer to this question has only been given for
$n=2\,;$ we will summarize it below.
We shall proceed, following the historical development, in two
steps.

First, we note that the full symmetry of the $2D\,$ zero mode
problem is given by
\be
\lb{fullsym}
{\cal U}_q = U_q(sl_2 )_\Delta \otimes U_q(sl_2 )_b
\ee
where the diagonal (coproduct) $U_q\,$ action $\Delta\,$ is
defined in (\ref{Deltainv}) (that is fully deciphered in Eqs. (2.2)
and (2.3) of Ref. \cite{DT2}) while the Chevalley generators of the
second factor are
\be
\lb{b}
b= a^1_\a {\bar a}^\a_2\,,\quad b' = -\; a^2_\a {\bar
a}^\a_1\quad{\rm and}\quad q^{\pm(p-{\bar p})}
\ee
satisfying
\be
\lb{b2}
[ b , b' ] = [ p  - {\bar p} ]\,,\quad
q^{p-{\bar p}}\; b\; =\; q^2 \;b\; q^{p-{\bar p}}\,,\quad
q^{p-{\bar p}}\; b'\; =\; \bq^2\; b'\; q^{p-{\bar p}}\,.
\ee
\vspace{3mm}

\noindent
{\bf Proposition 4.1}~(\cite{FHT3, DT})~{\em The subspace ${\cal
H}_I\,$ of ${\cal  U}_q$-invariant vectors of ${\cal H}\,$ is
$2h-1\,$ dimensional and is spanned by vectors of the form
\be
\lb{H_I}
|\lambda +1{\cal i}_I = {B'}^{~[\lambda ]}\, |1,0{\cal i}\otimes
|1,0{\cal i}\,,\quad
{B'}^{~[\lambda ]} = \sum\limits_{\nu =m}^{\lambda -m}
q^{\nu(\lambda -\nu )}\; {B'}_1^{~[\nu ]}\; {B'}_2^{~[\lambda-\nu ]}
\ee
where ${B'}_\a= a^1_\a\,{\bar a}^\a_1\,,\ \ \a = 1,2\,,\ \  m = \;
{\rm max}\ (0, \lambda -h+1)\,,\ \ 0\le\lambda\le2h-2\,.$ The
${\cal U}_q$-invariant subalgebra of observables of ${\cal A}_h\,$ is
generated by the pair $B' = B'_1 + B'_2\,,\
B = a^2_\a\,{\bar a}^\a_2 = B_1 + B_2\,$ satisfying
\be
\lb{obs}
B'\; B = [p]\;[p-1]\,,\quad
B \; B' = [p]\;[p+1]\quad (\;{\rm on}\ {\cal H}_I )\,;
\ee
the operators $B\,,\, B'\,,\, q^{\pm 2p}\,$ give rise to a $q$-deformation of the $su(1,1)\,$
Lie algebra:
\be
\lb{B}
[B , B' ] = [ 2p ]\,,\quad q^{2p} B = \bq^2 B\; q^{2p}\,,\quad
q^{2p} B' = q^2 B' q^{2p}\,.
\ee
$B\,$ and $B'\,$ act on the basis (\ref{H_I}) according to
\be
\lb{BonH_I}
B\; |p{\cal i}_I \;=\; [p]\, |p-1{\cal i}_I\,\quad
B'\; |p{\cal i}_I\; = \;[p]\, |p+1{\cal i}_I\,.
\ee
The restriction to ${\cal H}_I\,$ of the bilinear form on ${\cal
H}\,$ gives
\be
\lb{<>}
{\cal h} p | p'{\cal i} \; =\; \delta_{p p'} \;[p]\quad
(\ \Rightarrow\,\ {\rm sign}\;{\cal h} p | p{\cal i} \; =\;{\rm
sign}\;(h-p))\,.
\ee
}

The space ${\cal H}_I$ satisfies conditions (ii) and (iii) -- but
not (i) -- of the desiderata for the physical space ${\frak H}\,$
listed in the beginning. It was demonstrated in \cite{DT} that the
space of physical states can be obtained from ${\cal H}_I\,$ by
applying the generalized (BRS) cohomology of \cite{DVK, DV}.
\vspace{3mm}

\noindent
{\bf Theorem 4.2}~{\em

\noindent
(a) The conditions $B^h_\a = 0\,$ and $B_2 B_1 = q^2 B_1 B_2\,$
imply $B^h \equiv (B_1 +B_2 )^h = 0\,$ in ${\cal H}$.

\noindent
(b) Each of the generalized cohomologies
\be
\lb{gencohomol}
H^{(p)} ({\cal H}_I , B) \,=\, {\rm Ker} \;B^p \;/\; {\rm
Im}\;B^{h-p}\,,\quad p=1,\dots ,h-1
\ee
is one dimensional and is given by $\{\;{\Bbb C}\; |p{\cal i}\,, \
0<p<h\;\}\,.$
}
\vspace{3mm}

These results leave something to desire: in the above treatment 
the ${\cal U}_q\,$ symmetry of
\be
\lb{frakH}
{\frak H} = \oplus_{p=1}^{h-1} \; H^{(p)} ({\cal H}_I , B )
\ee
was not an outcome of the (generalized) BRS construction but a
precondition imposed on ${\cal H}_I \subset {\cal H}\,$. The first
question addressed (and answered in the negative) in \cite{DT2}
was: is there a $h$-differential complex $({\cal H} , Q )\,$ with
the same generalized cohomology as $({\cal H}_I , B )\,$?
\vspace{3mm}

\noindent
{\bf Proposition 4.3}~(\cite{DT2}, Section 3)~{\em
If $H^{(p)} ({\cal H} , Q) = H^{(p)} ({\cal H}_I , B)\,$ (for
$Q^h = 0 = B^h\,,\ \ 0<p<h\, )\,,$ then ${\rm dim}\,{\cal H} = \pm
1\ ( {\rm mod}\, h )\,.$
}
\vspace{3mm}

This result excludes the existence of a generalized BRS charge
$Q\,$ in ${\cal H}\,$ with the desired properties since ${\rm
dim}\ {\cal H}\; = h^4\,$ for ${\cal H}\,$ given by (\ref{findimstate}).

A solution to the resulting puzzle, given in \cite{DT2} (Section
4), consists in embedding ${\cal H}\,$ in a graded vector space
\be
\lb{bullet}
{\cal H}^\bullet = \oplus_{\nu\ge 0}\;{\cal H}^\nu\ {\rm with}\
{\cal H}^0 = {\cal H}\,, \quad {\cal H}^\nu = {\cal H} / {\cal H}_I\,,\
1\le\nu\le h-1\,,\quad {\cal H}^\nu = 0\,,\ \nu\ge h\,,
\ee
introducing a $h$-differential, $d :\; {\cal H}^\nu\;\rightarrow
{\cal H}^{\nu +1}\ (d^h = 0)\,$ and extending $B\,$ to 
${\cal H}^\bullet\,$ in such a way that the nilpotent operator 
$Q = d+B\,$ defines the same cohomology in ${\cal H}^\bullet\,$ as
$B\,$ in ${\cal H}_I\,$ (Theorem 1 of \cite{DT2}):
\be
\lb{Q^h}
Q^h\equiv (d+B)^h = 0\,,\quad H^{(p)} ({\cal H}^\bullet , Q ) =
H^{(p)} ({\cal H}_I , B )\,.
\ee
A "physical geometric" interpretation of this result in terms of
(generalized) Hochschild cochains was also proposed in \cite{DT2}
(Sections 5,6).

\section{Concluding remarks}

The finite dimensional gauge problem extracted from the WZNW model
displays a rich structure and opens the way of applying recently
developed generalizations of BRS cohomology. It supports the
maverick opinion that, contrary to the common beliefs, the true
understanding of solvable $2D\,$ current algebra models still lies
in the future.

The results of Section 4 only apply to the $SU(2)\,$ WZNW model.
Their extension to higher rank groups offers an obvious suggestion
for further work. More important, in our view, is the problem of
making full use of the extended space, the necessity of which is
indicated in Proposition 4.3. The possibility to relate it to  the recently advanced operad approach to quantum
field theory \cite{Kon} looks tantalizing.

\section*{Acknowledgements}
One of the authors (I.T.T.) would like to thank Professor H.-D.
Doebner and the organizers of the Symposium "Quantum Theory and
Symmetries" for their hospitality in Goslar where this paper was
presented.
The final version of this report is been completed while M.D.-V. and
I.T.T. are visiting ESI, Vienna, and L.K.H. is visiting ICTP, Trieste.
The authors thank these institutions for hospitality. 
P.F. acknowledges the support of the Italian Ministry of University, Scientific
Research and Technology (MURST).
This work is supported in part by the Bulgarian National Council
for Scientific Research under contract F-828, and by CNRS and RFBR grants
PICS-608, RFBR 98-01-22033.


\begin{thebibliography}{**}

\bibitem{HIOPT}
L.K. Hadjiivanov, A.P. Isaev, O.V. Ogievetsky,
P.N. Pyatov, I.T. Todorov. 
Hecke algebraic properties of dynamical
$R$-matrices. Application to related matrix algebras.
q-alg/9712026; 
{\em J. Math. Phys.} {\bf 40} (1999) 427-448. 

\bibitem{W}
E. Witten.
Non-abelian bosonization in two dimensions. 
{\em Commun. Math. Phys.} {\bf 92} (1984) 455-472.

\bibitem{KZ}
V.G. Knizhnik, A.B. Zamolodchikov.
Current algebra and Wess-Zumino model in two dimensions. 
{\em Nucl. Phys.} {\bf B247} (1984) 83-103.\\
I.T. Todorov.
Infinite Lie algebras in $2$-dimensional conformal field theory.
Proceedings of the XIII International conference on 
Differential Geometric
Methods in Physics, Shumen, Bulgaria 1984, H.-D. Doebner, T.
Palev eds., World Scientific, Singapore 1986, pp. 297-347.\\
Current algebra approach to conformal invariant two-dimensional models.
{\em Phys. Lett.} {\bf B153} (1985) 77-81.

\bibitem{BPZ}
A.A. Belavin, A.M. Polyakov, A.B. Zamolodchikov.
Infinite conformal symmetry in two-dimensional quantum field theory.
{\em Nucl. Phys.} {\bf B241} (1984) 333-380.\\
A.B. Zamolodchikov, V.A. Fateev.
Operator algebra and correlation functions in the two-dimensional
$SU(2)\times SU(2)\,$ chiral Wess-Zumino model.
{\em Yad. Fiz.} {\bf 43} (1986) 1031-1044 (English translation: {\em Sov. 
Journ. Nucl. Phys.} {\bf 43:4} (1986) 657-664).

\bibitem{V}
E. Verlinde.
Fusion rules and modular transformations in $2D\,$ conformal field theory. 
{\em Nucl. Phys.} {\bf B300} (1988) 360-376.

\bibitem{B}
O. Babelon. 
Extended conformal algebra and the Yang-Baxter equation.
{\em Phys. Lett.} {\bf B215} (1988) 523-529.\\
B. Blok. 
Classical exchange algebras in the Wess-Zumino-Witten model.
{\em Phys. Lett.} {\bf B233} (1989) 359-362.

\bibitem{F1}
L.D. Faddeev.
On the exchange matrix for WZNW model. 
{\em Commun. Math. Phys.} {\bf 132} (1990) 131-138.\\
A. Alekseev, S. Shatashvili. 
Quantum groups and WZNW models.
{\em Commun. Math. Phys.} {\bf 133} (1990) 353-368.

\bibitem{F2}
L.D. Faddeev. 
Quantum symmetry in conformal field theory by Hamiltonian methods, 
Garg\`ese 1991, 
{\em in}
New Symmetry Principles in Quantum Field
Theory, J. Fr\"ohlich et al. eds., Plenum Press, N.Y. 1992,
pp. 159-175.\\
A.Yu. Alekseev, L.D. Faddeev, M.A. Semenov-Tian-Shansky.
Hidden quantum groups inside Kac-Moody algebras.
{\em Commun. Math. Phys.} {\bf 149} (1992) 335-345.

\bibitem{G}
K. Gaw\c{e}dzki. 
Classical origin of quantum group symmetries in Wess-Zumino-Witten
conformal field theory. 
{\em Commun. Math. Phys.} {\bf 139} (1991) 201-213.

\bibitem{FG}
F. Falceto, K. Gaw\c{e}dzki. 
Lattice Wess-Zumino-Witten model and quantum groups. 
{\em J. Geom. Phys.} {\bf 11} (1993) 251-279.

\bibitem{CGHOS}
M. Chu, P. Goddard, I. Halliday, D. Olive, A. Schwimmer.
Quantization of the Wess-Zumino-Witten model on a circle.
{\em Phys. Lett.} {\bf B266} (1991) 71-81.\\
M. Chu, P. Goddard. 
Quantisation of the $SU(N)\,$ WZW model at level $k\,.$ hep-th/9407116; 
{\em Nucl. Phys.} {\bf B445} (1995) 145-168.

\bibitem{FHT1}
P. Furlan, L.K. Hadjiivanov, I.T. Todorov.
Canonical approach to the quantum WZNW model.
Trieste-Vienna preprint IC/95/74; ESI 234 (1995).

\bibitem{FHT2}
P.Furlan, L.K.Hadjiivanov, I.T. Todorov.
Operator realization of the $SU(2)\,$ WZNW model.
hep-th/9602101; 
{\em Nucl. Phys.} {\bf B474} (1996) 497-511.

\bibitem{FHT3}
P. Furlan, L.K. Hadjiivanov, I.T. Todorov.
A quantum gauge group approach to the $2D\ SU(n)\,$ WZNW model.
hep-th/9610202;
{\em Int. J. Mod. Phys.} {\bf A12} (1997) 23-32.

\bibitem{AF}
A.Yu. Alekseev, L.D. Faddeev.
$(T^* G)_t :$ A toy model for conformal field theory. 
{\em Commun. Math. Phys.} {\bf 141} (1991) 413-422.

\bibitem{BF}
A.G. Bytsko, L.D. Faddeev.
$(T^* B)_q\,,\ q$-analogue of model space and CGC generating matrices. 
q-alg/9508022;
{\em J. Math. Phys.} {\bf 37} (1996) 6324-6348.

\bibitem{AFFS}
A.Yu. Alekseev, L.D. Faddeev, J. Fr\"ohlich, V. Schomerus.
Representation theory of lattice current algebras. q-alg/9604017; 
{\em Commun. Math. Phys.} {\bf 191} (1998) 31-60.

\bibitem{BS}
A.G. Bytsko, V. Schomerus.
Vertex operators -- from a toy model to lattice algebras. 
q-alg/9611010;
{\em Commun. Math. Phys.} {\bf 191} (1998) 87-136.

\bibitem{DT}
M. Dubois-Violette, I.T. Todorov.
Generalized cohomologies and the physical subspace of the $SU(2)\,$ WZNW
model. 
hep-th/9704069; 
{\em Lett. Math. Phys.} {\bf 42} (1997) 183-192. 

\bibitem{DT2}
M. Dubois-Violette, I.T. Todorov.
Generalized homologies for the zero modes of the $SU(2)\,$ WZNW model. 
math.QA/9905071;
{\em Lett. Math. Phys.} {\bf 48} (1999) 323-338.

\bibitem{BRS}
C. Becchi, A. Rouet, R. Stora.
Renormalization models with broken symmetries, 
{\em in} Renormalization Theory, Erice 1975, G. Velo, A.S.
Wightman eds., Reidel 1976, pp. 299-343.\\
Renormalizable theories with symmetry breaking, 
{\em in} Field Theory,
Quantization and Statistical Mechanics, E. Tirapegui ed.,
Reidel 1981, pp. 3-32.

\bibitem{FHIPT}
P. Furlan, L.K. Hadjiivanov, A.P. Isaev,
P.N. Pyatov, I.T. Todorov.
Quantum matrix algebra for the $SU(n)$ WZNW model 
(in preparation).

\bibitem{BFP}
J. Balog, L. Feh\'er, L. Palla. 
The chiral WZNW phase space and its Poisson-Lie groupoid.
hep-th/9907050; 
{\em Phys. Lett.}  {\bf B 463} (1999) 83-92.\\ 
Chiral extensions of the WZNW phase space,
Poisson-Lie symmetries and groupoids. hep-th/9910046. 

\bibitem{S-T-S}
M.A. Semenov-Tian-Shansky.
Dressing transformations and Poisson group actions. 
{\em Publ. RIMS, Kyoto Univ.} {\bf 21} (1985) 1237-1260.

\bibitem{AT}
A.Yu. Alekseev, I.T. Todorov.
Quadratic brackets from symplectic forms. hep-th/9307026; 
{\em Nucl.Phys.} {\bf B421} (1994) 413-428.

\bibitem{FRT}
L.D. Faddeev, N. Yu. Reshetikhin, L.A. Takhtajan.
Quantization of Lie groups and Lie algebras. 
{\em Algebra i Analiz} {\bf 1:1} (1989) 178-206 
(English translation: {\em Leningrad Math. J.} {\bf 1} (1990) 193-225).

\bibitem{Dr}
V.G. Drinfeld.
Quantum groups, 
{\em in} Proceedings of the International Congress of
Mathematicians, Berkeley 1986, Academic Press, N.Y. 1986, v.1,
pp. 798-820.

\bibitem{Jimbo}
M. Jimbo.
A $q$-analogue of $U({\frak g}{\frak l}(N+1))\,,$ Hecke algebra, and the
Yang-Baxter equation.
{\em Lett. Math. Phys.} {\bf 11} (1986) 247-252.

\bibitem{Ful}
W. Fulton.
Young Tableaux: With Applications to Representation Theory
and Geometry,
Cambridge Univ. Press, 1997.

\bibitem{CP}
V. Chari, A. Pressley.
A Guide to Quantum Groups, 
Cambridge Univ. Press, 1994.

\bibitem{GW}
D. Gepner, E. Witten.
String theory on group manifolds.
{\em Nucl. Phys.} {\bf B278} (1986) 493-549. 

\bibitem{TK}
A. Tsuchiya, Y. Kanie.
Vertex operators in the conformal field theory on ${\Bbb P}^1\,$ and
monodromy representations of the braid group.
{\em Lett. Math. Phys.} {\bf 13} (1987) 303-312.

\bibitem{STH}
Y.S. Stanev, I.T. Todorov, L.K. Hadjiivanov.
Braid invariant rational conformal models with a quantum group symmetry.
{\em Phys. Lett.} {\bf B276} (1992) 87-94. 

\bibitem{BBB}
V. Pasquier. 
Etiology of IRF models. 
{\em Commun. Math. Phys.} {\bf 118} (1988) 355-364.\\
O. Babelon, D. Bernard, E. Billey. 
A quasi-Hopf algebra interpretation of quantum 3-$j\,$ and 6-$j\,$
symbols and difference equations. 
q-alg/9511019;
{\em Phys. Lett.} {\bf B375} (1996) 89-97.\\
O. Babelon.
Universal exchange algebra for Bloch waves and Liouville theory. 
{\em Commun. Math. Phys.} {\bf 139 } (1991) 619-643.

\bibitem{GN}
J.-L. Gervais, A. Neveu.
Novel triangle relation and absence of tachyons in Liouville string field
theory. 
{\em Nucl. Phys.} {\bf B238} (1984) 125-141.

\bibitem{I}
A.P. Isaev.
Twisted Yang-Baxter equations for linear quantum (super)groups.
q-alg/9511006; 
J. Phys. A: Math. Gen. {\bf 29} (1996) 6903-6910.

\bibitem{D2}
V.G. Drinfeld.
Quasi-Hopf algebras.
{\em Algebra i Analiz} {\bf 1:6} (1989) 114-148 (English translation:
{\em Leningrad Math. J.} {\bf 1} (1990) 1419-1457).\\
N.Yu. Reshetikhin.
Multiparameter quantum groups and twisted quasitriangular Hopf algebras.
{\em Lett. Math. Phys.} {\bf 20} (1990) 331-335.

\bibitem{L}
G. Lusztig. 
Canonical bases arising from quantized enveloping algebras.
{\em Journ. Amer. Math. Soc.} {\bf 3} (1990) 447-498.\\
Introduction to Quantum Groups, Progress in Math. {\bf 110}, Birkh\"auser,
Boston-Basel-Stuttgart 1993. 

\bibitem{DVK}
M. Dubois-Violette, R. Kerner.
Universal $q$-differential calculus and
$q$-analog of homological algebra.
{\em Acta Math. Univ. Comenian.} {\bf 65} (1996) 175-188. 

\bibitem{DV}
M. Dubois-Violette.
$d^N=0 :\;$ Generalized homology. 
{\em K-Theory} {\bf 14} (1998) 371-404.

\bibitem{Kon}
M. Kontsevich. Operads and motives in deformation quantization. 
math/ 9904055; {\em Lett. Math. Phys.} {\bf 48} (1999) 35-72.

\end{thebibliography}
\end{document}